\documentclass{style/nseJournal}

\usepackage{tikz}
\usetikzlibrary{arrows.meta, positioning, shapes.geometric, decorations.pathmorphing, fadings, calc}

\usepackage{booktabs}
\usepackage[Symbol]{upgreek}

\begin{document}

\title{Photonuclear Neutron Production in OpenMC: Verification Against MCNPX, FLUKA, and a First-Collision Analytical Solution}

% Use the \addAuthor macro to add authors in the order they should appear. The second argument corresponds to
% the affiliation dec¢lared below.
% The corresponding author should be wrapped in \correspondingAuthor

\addAuthor{Lorenzo Loi}{a}
\addAuthor{Andrea Missaglia}{e,a}
\addAuthor{Guy Stein}{f,g}
\addAuthor{Carolina Introini}{a}
\addAuthor{David Breitenmoser}{d}
\addAuthor{Shaun D. Clarke}{d}
\addAuthor{Sara A. Pozzi}{d}
\addAuthor{\correspondingAuthor{Antonio Cammi}}{a,b,c}
% The corresponding author's email can be specified using \correspondingEmail
\correspondingEmail{antonio.cammi@polimi.it}

%%% Affiliations (from authblk)
%%% \addAffiliation{affiliationNumber}{Name of Institute, City, State/Country}
\addAffiliation{a}{Politecnico di Milano, Department of Energy,\\ Via La Masa 34, Milano 20133, Italy}
\addAffiliation{b}{Department of Mechanical and Nuclear Engineering, Khalifa University,\\ Abu Dhabi, 127788, United Arab Emirates}
\addAffiliation{c}{Emirates Nuclear Technology Centre (ENTC),\\ Khalifa University of Science and Technology, \\Abu Dhabi, 127788, United Arab Emirates}
\addAffiliation{d}{Department of Nuclear Engineering and Radiological Sciences, \\University of Michigan, Ann Arbor, MI 48109-2104, USA}

\addAffiliation{e}{Department of Nuclear Engineering, Texas A\&M University,\\ 423 Spence St, College Station, 77843-3133, TX, USA}
\addAffiliation{f}{School of Physics and Astronomy, Tel Aviv University, Tel Aviv 6997801, Israel}
\addAffiliation{g}{Nuclear Research Centre Negev, P.O. Box 9001, Beer-Sheva, Israel}

% Add keywords to appear in Abstract in the order they should appear
\addKeyword{photonuclear reactions}
\addKeyword{OpenMC}
\addKeyword{Monte Carlo verification}
\addKeyword{photoneutron production}
\addKeyword{MCNPX}
\addKeyword{FLUKA}
\addKeyword{photonuclear data}
\addKeyword{broomstick benchmark}

\date{\vspace{1em} \small }

\titlePage

\begin{abstract}
The modelling of photonuclear reactions is increasingly important for applications involving high-energy photon fields, including accelerator-driven neutron sources, radiation shielding, medical physics, and fusion technologies. Although OpenMC provides well-verified photoatomic transport capabilities, its photonuclear physics is currently available only in an unofficial development branch and requires independent verification before broader scientific use. This work presents a systematic verification of the OpenMC photonuclear implementation using six single-collision broomstick numerical benchmarks based on $^{2}$H, $^{9}$Be, and $^{238}$U targets, irradiated by monoenergetic 5~MeV and 15~MeV photons and by a continuous 1--20~MeV LINAC-representative spectrum. OpenMC was compared with MCNPX under common ENDF7u photonuclear data, with FLUKA as an independent, native-model reference, and with a first-collision analytical solution likewise based on ENDF7u; calculations with the IAEA/PD-2019 library further quantified nuclear-data sensitivity. Energy-resolved, angular, and integrated photoneutron yields were compared across all codes and evaluations. Under common ENDF7u data, OpenMC and MCNPX agreed to within a few tenths of a percent in integrated yield, consistent with the statistical uncertainty of the calculations. Their energy and angular distributions also agreed closely, apart from small kinematic differences in the light-nuclide cases. Replacing ENDF7u with IAEA/PD-2019 produced deviations from OpenMC–ENDF7u of up to 13.9\%, largest for continuous-source $^9$Be and monoenergetic $^2$H (11.3\%). FLUKA differed from OpenMC–ENDF7u by up to $\sim$9\% (monoenergetic) and $\sim$2.5\% (continuous-source) and predicted a markedly different angular distribution for $^2$H than the evaluated-data calculations. These results verify that OpenMC's photonuclear implementation reliably reproduces MCNPX under a given nuclear-data evaluation. Which evaluation is itself most accurate, however, remains open: ENDF7u, IAEA/PD-2019, and FLUKA's native models agree closely in some cases and diverge in others, so no single library or model is uniformly preferable from the present comparisons alone.
\end{abstract}

\newpage
\section{Introduction}

Photon-induced nuclear reactions are relevant to a wide range of scientific and engineering applications, including photon-driven subcritical systems, medical isotope production, radiation shielding, activation analyses, radiotherapy, and the design of high-energy accelerator facilities~\cite{ZILGES2022103903,Gohar_OSTI,TSECHANSKI2016124,UDDIN2024101031}. When the incident photon energy exceeds the separation threshold of a target nucleus, neutron-emitting reactions such as $(\upgamma,\text{n})$ and $(\upgamma,\text{2n})$ may occur, together with photofission in sufficiently heavy nuclei. In the energy range relevant to most accelerator-driven applications, photonuclear cross sections are largely governed by the Giant Dipole Resonance, a collective nuclear excitation generally extending over the few-to-tens-of-MeV range~\cite{Findlay_1990,DietrichBerman1988,Filipescu2024}. Reliable prediction of the resulting neutron yields, energy spectra, angular distributions, and reaction-channel contributions is therefore essential whenever high-energy photon fields (i.e., $\geq$ 1~MeV) interact with matter.

Monte Carlo particle-transport codes are the principal computational tools used to model these systems; however, their predictive reliability cannot be inferred solely from the physical completeness of the implemented models or from the availability of evaluated nuclear data. Following the terminology of the ASME V\&V standards~\cite{ASME_VV10_2019,ASME_VV_20_2009}, verification is the process of determining that a computational model accurately represents the underlying mathematical model and its solution, whereas validation determines the degree to which the model is an accurate representation of the real world from the perspective of its intended uses. Verification is further divided into code verification, which establishes that the mathematical model is correctly implemented in the code, preferably through comparison with analytical solutions and, less rigorously, with established reference codes, and solution verification, which estimates the numerical error of a given calculation, here the Monte Carlo statistical uncertainty. A new transport capability must first undergo code verification under controlled conditions before it can be validated against experimental observations; the present work addresses this first step.

OpenMC is an open-source, continuous-energy Monte Carlo transport code whose neutron and photoatomic capabilities have been extensively developed and applied~\cite{OpenMC_Romano}. Photon transport was introduced by Lund \textit{et al.}~\cite{Lund2018Implementation}, including coherent and incoherent scattering, photoelectric absorption, pair production, atomic relaxation, and an approximate treatment of secondary-electron bremsstrahlung. Subsequent studies have exercised these capabilities in cross-code comparisons involving photon transport, shutdown-dose calculations, deposited energy, and coupled neutron--photon applications~\cite{ZHANG2020111837,Bae_2022,Peterson2024Development}. This body of work provides substantial evidence for the reliability of the photoatomic physics distributed with official OpenMC releases.\\
The status of photonuclear physics is fundamentally different. At the time of this study, photoneutron reactions were not included in the official OpenMC distribution and remained explicitly identified as an unavailable feature in the official documentation~\cite{OpenMCDocsPhoton}. The capability investigated here was instead implemented in an independent development branch maintained by Guy Stein~\cite{SteinRepo} (version 0.16.1-dev138). Consequently, it had not undergone the verification and review process normally associated with the integration of a new physics capability into an official release.  Photonuclear verification and validation studies remain comparatively limited in the broader Monte Carlo literature. The Barber and George benchmark provides a reference framework for testing photonuclear transport, while Tuyet \textit{et al.} compared photonuclear yields and secondary-particle spectra calculated with TRIPOLI-4, DIANE, and MCNP6~\cite{BarberGeorge,Tuyet2023Comparison}. These studies demonstrate that agreement in total neutron production does not necessarily imply agreement in the energy or angular distributions of the emitted particles. To the authors' knowledge, prior to the preliminary conference contribution associated with the present work~\cite{LOI_OpenMC_verification}, no peer-reviewed verification had specifically examined the photonuclear implementation contained in the unofficial OpenMC development branch considered here.

A further difficulty is introduced by the photonuclear data themselves. Sari~\cite{Sari2023} identified substantial deficiencies in evaluated photonuclear libraries and their use in common transport codes, including inaccurate reaction thresholds, incomplete or inconsistent cross sections, and errors affecting secondary-neutron kinematics in the 4--20~MeV range. These issues can produce significant discrepancies between calculated and experimental photoneutron yields, up to factors 2--3 when compared with experimental measurements. Verification of a transport implementation must, therefore, separate as far as possible, differences caused by the code from those originating in the underlying evaluation. In the present study, OpenMC and MCNPX are first compared using the same ENDF7u photonuclear data derived from ENDF/B-VII.1~\cite{ENDFVII}, providing a direct test of the new OpenMC implementation under common nuclear data. FLUKA is then included as a complementary reference based on its native photonuclear models rather than on the same evaluated photonuclear library. Finally, OpenMC calculations performed with the more recent IAEA/PD-2019\footnote{Evaluations with JENDL-5 nuclear data have also been performed, showing consistent results with IAEA/PD-2019.} evaluation~\cite{Kawano2020IAEA} are used to quantify the sensitivity of the results to the selected photonuclear data. 

A previous study~\cite{LOI_OpenMC_verification} considered the canonical \textit{broomstick}, or \textit{pencil-beam}, numerical benchmark~\cite{broomstick} for three monoenergetic photon sources incident on $^{2}$H, $^{9}$Be, and $^{238}$U targets. The extremely small transverse dimension of the target suppresses multiple interactions and secondary transport, providing a controlled environment in which photonuclear reaction probabilities and emitted-neutron spectra can be examined with minimal interference from macroscopic geometry effects. The three materials were selected to exercise substantially different physical regimes: two-body deuteron photo-disintegration, low-probability multi-body breakup in $^{9}$Be, and the competition among $(\upgamma,\text{n})$, $(\upgamma,\text{2n})$, and photofission channels in $^{238}$U.

This work substantially extends that preliminary assessment and, accordingly, aims to verify the photonuclear implementation of the unofficial OpenMC development branch against MCNPX under common evaluated nuclear data, and to assess its behavior against the independent native photonuclear treatment available in FLUKA. First, the three monoenergetic numerical benchmarks are complemented by three calculations using a continuous, linear accelerator (LINAC)-representative photon distribution based on the Kramers bremsstrahlung form~\cite{Kramers1923}, testing whether code differences observed at individual photon energies persist or compensate when folded over a broad source spectrum. Second, updated FLUKA calculations are included for all six configurations, providing an independent model-based reference in addition to the common-data OpenMC--MCNPX comparison. Third, the calculated integrated neutron yields are compared with an independent first-collision analytical solution based on the competition between photoatomic and neutron-producing photonuclear interactions~\cite{GRYAZNYKH_2000}. Fourth, the influence of replacing ENDF7u with IAEA/PD-2019 in OpenMC is evaluated explicitly to quantify the sensitivity of the predictions to the underlying photonuclear evaluation. Fifth, the angular distribution of the emitted neutron current is compared across codes and evaluations in addition to the energy distributions, allowing agreement in total production to be distinguished from agreement in reaction kinematics and angular emission patterns. The study is deliberately restricted to verification under idealized, single-collision conditions; comparison with experimental photonuclear measurements, and therefore formal validation of the OpenMC capability, is reserved for subsequent work.

The remainder of the paper is organized as follows. Section~\ref{sec:methods}  describes the broomstick numerical benchmark geometry, target materials, photon sources, nuclear-data libraries, Monte Carlo implementations, scoring procedures, and first-collision analytical solution. Section~\ref{sec:results} presents the energy-resolved, angular, and integrated neutron-yield comparisons, while Section~\ref{sec:discussion} examines the implications of the common-data verification, the differences between integral and differential (spectral and angular) agreement, the sensitivity to photonuclear data, and the interpretation of the FLUKA comparison. Finally, Section~\ref{sec:conclusions} summarizes the main findings, limitations, and requirements for future experimental validation.

\section{Methods}
\label{sec:methods}

\subsection{Broomstick benchmark and case definition}
\label{sec:benchmark}

\begin{figure}
    \centering
    \includegraphics[width=1\linewidth]{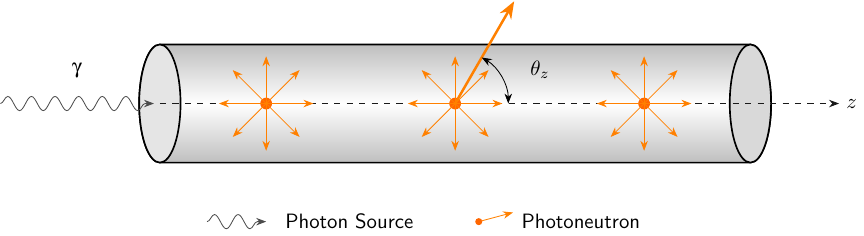}
    \caption{Schematic representation of the photonuclear broomstick benchmark. A mono-directional photon source is aligned with an extremely thin cylindrical target surrounded by vacuum. Photonuclear interactions produce secondary neutrons that escape predominantly through the lateral target surface. The radial dimension is not shown to scale.}
    \label{fig:broom_scheme}
\end{figure}

The \textit{broomstick}, or \textit{pencil-beam}, problem is a canonical verification numerical benchmark designed to isolate the treatment of individual particle interactions from secondary transport and macroscopic geometrical effects~\cite{broomstick}. In its general computational formulation, a mono-directional source is aligned with the axis of an extremely long and narrow cylindrical target surrounded by vacuum. The dimensions are selected so that the target is effectively optically thick in the axial direction and optically thin in the transverse direction. For example, Cullen et al.~\cite{Cullen2003Broomstick} employed a cylinder $10^{5}$~cm long and $10^{-8}$~cm in radius to ensure that nearly every source particle underwent a collision along the cylinder axis, while particles emerging from that collision escaped through the lateral surface without appreciable secondary transport.\\
In the present work, the numerical benchmark was adapted to photonuclear neutron production. As shown in Figure~\ref{fig:broom_scheme}, a photon source located at $z=0$ was directed along the positive $z$-axis toward a cylindrical target. The target radius was set to $1\times10^{-6}$~cm across all configurations. This specific dimension acts as an optimal lower limit for a one-dimensional approximation; further reduction introduces severe floating-point precision errors that degrade tracking reliability and result in prohibitive computational costs. The target length was selected separately for each material, following the reference MCNPX input decks, so that the probability of a primary photon reaching the downstream boundary without undergoing an interaction was negligible. The adopted target lengths were 3000~cm for $^{2}$H, 1500~cm for $^{9}$Be, and 400~cm for $^{238}$U. Because the upstream target surface was located at $z=2$~cm, the corresponding downstream boundaries were at $z=3002$, 1502, and 402~cm, respectively. These exact dimensions supersede the rounded common target length reported in the preliminary conference contribution~\cite{LOI_OpenMC_verification}.

The target and source definitions were consistently reproduced in OpenMC, MCNPX, and FLUKA. The geometry is intentionally non-physical from an engineering perspective: its purpose is not to represent a practical photoneutron converter, but to provide a controlled single-collision limit in which differences among evaluated data, interaction sampling, and secondary-particle kinematics can be identified without substantial interference from transport effects.

\begin{figure}[htbp]
    \centering
    \includegraphics[width=0.62\linewidth]{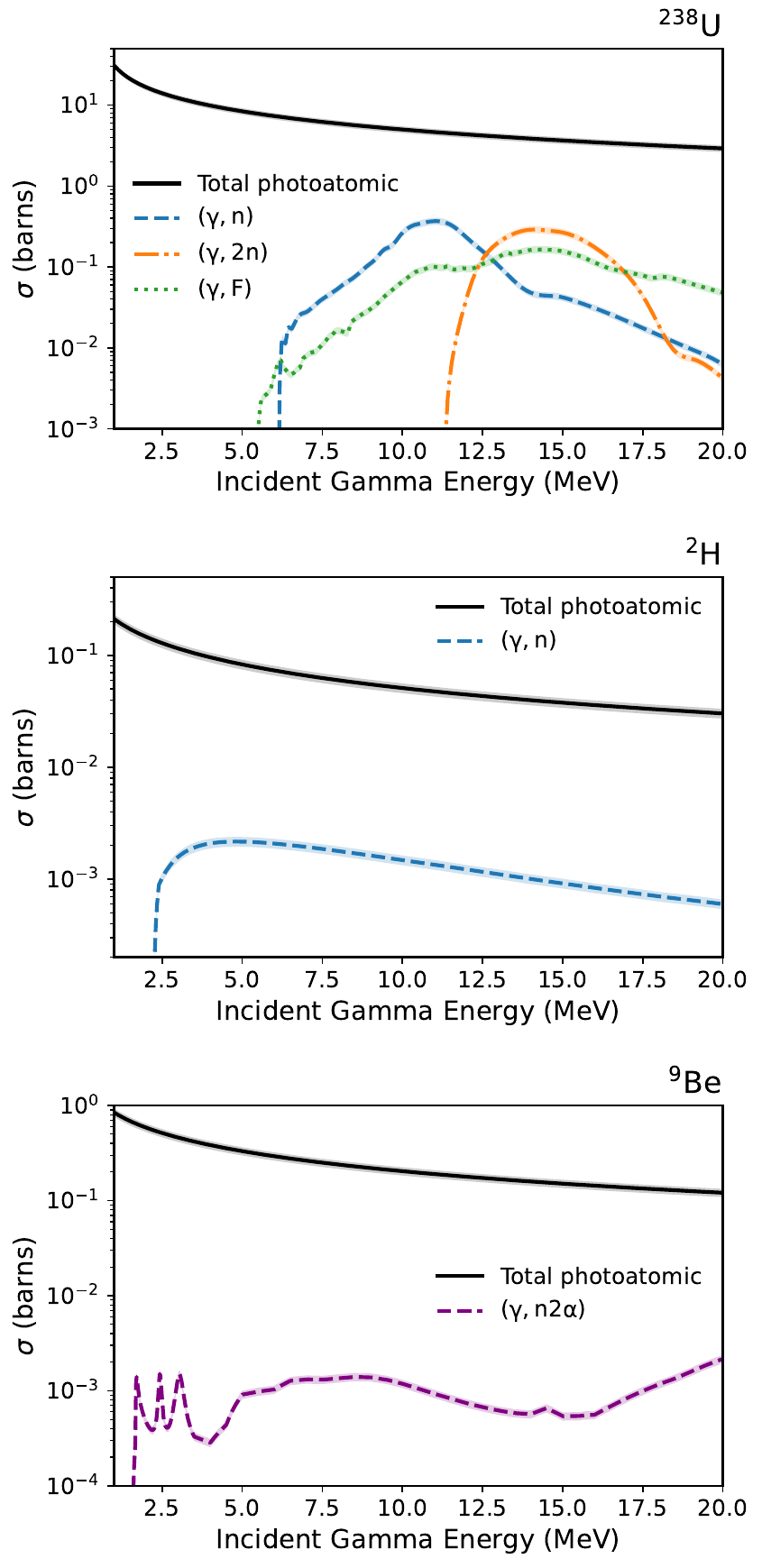}
    \caption{Microscopic photoatomic and neutron-producing photonuclear cross sections for $^{2}$H, $^{9}$Be, and $^{238}$U over the photon-energy range relevant to this study. The photonuclear curves correspond to the ENDF7u evaluation used by OpenMC and MCNPX. FLUKA photonuclear interactions were instead treated using its native models.}
    \label{fig:xs}
\end{figure}

Three isotopically pure target materials were selected to exercise distinct photonuclear regimes. Deuterium represents a low-threshold, predominantly two-body photodisintegration case governed by the $^{2}$H$(\upgamma,\text{n})$ reaction. Beryllium-9 provides a low-cross-section multi-body breakup case, dominated by the $^{9}$Be$(\upgamma,\text{n}2\upalpha)$ reaction. Uranium-238 represents a heavy-nuclide, multi-channel case in which $(\upgamma,\text{n})$, $(\upgamma,\text{2n})$, and photofission contribute simultaneously. The monoenergetic photon energies were 5~MeV for $^{2}$H and $^{9}$Be and 15~MeV for $^{238}$U. The choice of these particular energies was motivated by selecting a region where the cross sections were far from the lower threshold or from resonances (see Fig.~\ref{fig:xs}). Each target was also irradiated using the continuous, LINAC-representative source described in Section~\ref{sec:sources}.\\
The exact material and geometric definitions are summarized in Table~\ref{tab:benchmark_geometry}. Atomic densities were used directly in MCNPX and OpenMC. For FLUKA, they were converted into the corresponding mass densities while preserving the same isotopic number densities.

\begin{table}[htbp]
\centering
    \small
    \caption{Exact target definitions adopted in the six photonuclear broomstick scenarios.}
    \begin{tabular}{lcccccc}
        \toprule
        \textbf{Nuclide}

        & \textbf{Atomic density}
        
        & \textbf{$E_{\upgamma,\mathrm{mono}}$} & \textbf{Length}\\

        & \textbf{atom/(barn cm)}

        & \textbf{(MeV)} & \textbf{(cm)}\\
        \midrule
        $^{2}$H

        & 0.1806

        & 5  & 3000\\
        $^{9}$Be

        & 0.066822
        
        & 5 & 1500\\
        $^{238}$U

        & 0.0047433

        & 15 & 400\\
        \bottomrule
    \end{tabular}
    \label{tab:benchmark_geometry}
\end{table}

Figure~\ref{fig:xs} shows the photoatomic and neutron-producing photonuclear cross sections used to interpret the selected cases. The plotted photonuclear data correspond to the ENDF7u evaluation adopted in the common-data OpenMC, MCNPX and analytical comparison. They do not represent the native photonuclear interaction model used by FLUKA.

\subsection{Monte Carlo implementations and nuclear data}
\label{sec:mc_implementations}

The numerical benchmark was independently implemented in OpenMC, MCNPX, and FLUKA. The OpenMC calculations were performed using the unofficial photonuclear-development branch maintained by Guy Stein~\cite{SteinRepo} (version 0.16.1-dev138). ENDF7u photonuclear data were used for the principal OpenMC calculations, and photonuclear interaction biasing was applied to improve the sampling efficiency of the relatively rare neutron-producing events. Each Monte Carlo configuration used $10^{8}$ source photons.\\

The secondary energy-angle representation adopted for each photonuclear reaction channel and evaluated library was inspected directly from the processed nuclear-data files using the \texttt{openmc.data.IncidentPhotonuclear} interface, which exposes the underlying \texttt{AngleEnergy} distribution (e.g. Kalbach-Mann, correlated, or uncorrelated angle-energy laws) associated with each reaction. This inspection, summarized in Section~\ref{sec:discussion}, was used to interpret the origin of the angular differences observed between codes and evaluations rather than to alter the transport calculations themselves.

The reference evaluated-data calculations were performed using MCNPX 2.7.0~\cite{MCNPX270} due to compatibility with previous analysis~\cite{LOI_OpenMC_verification}. To confirm that this choice does not affect the generality of the verification, all six scenarios were additionally repeated with MCNP6.2~\cite{MCNP6} using the same ENDF7u photonuclear data, geometry, and source definitions. As discussed in Appendix~\ref{appendixA}, the two codes produced equivalent results in all six scenarios, so the conclusions of this work do not depend on the MCNP version adopted.
Photons and neutrons were included in the particle-transport mode, while an electron cutoff of 20~MeV effectively suppressed secondary-electron transport over the energy range considered. Surface-current estimators were used for the escaping neutron distributions.\\
The independent model-based calculations were performed using FLUKA version 4-5.0 with the PRECISIO default settings~\cite{FLUKA2014,FLUKA2015}. Pointwise neutron transport below 20 MeV was adopted, with the ENDF/B-VII.1 library explicitly selected through LOW-PWXS, while photonuclear interactions were treated using the native FLUKA models activated by PHOTONUC. The electron cutoff was chosen consistently with the other Monte Carlo references, and recoil protons were discarded from transport. For the $^{238}$U cases, coalescence and heavy-fragment evaporation were enabled. The resulting verification hierarchy is consequently as follows: OpenMC and MCNPX provide the primary code-implementation comparison under common ENDF7u photonuclear data, whereas FLUKA provides a complementary comparison involving an independent photonuclear model. Differences involving FLUKA may therefore reflect both code implementation and the underlying photonuclear interaction treatment.\\

The primary verification observable was the one-way outward neutron current crossing the cylindrical lateral surface of the target, $J_n$. All reported currents were normalized per source photon (i.e., a unit scoring area was considered). Under the single-collision assumptions of the broomstick geometry described in Section~\ref{sec:benchmark}, secondary
transport and re-entrant contributions are negligible, so that the scored outward current coincides, to the accuracy of the present cases, with the neutron production yield per source photon. Consistent with this equivalence, and with the notation already adopted for the Monte Carlo tallies throughout this work, the scored quantity is denoted $Y_n$ in the remainder of the paper. The reported bin values should therefore be interpreted as neutron crossings per source photon within each energy interval, rather than as spatially averaged fluences over the physical cylinder surface. 

In addition to the energy-resolved yield, the angular distribution of the escaping neutron current was scored as a function of the polar angle $\theta_z$ between the outward neutron direction and the incident photon axis (Figure~\ref{fig:broom_scheme}). By this definition, $\theta_z=0$ ($\mu=1$) corresponds to emission in the forward direction of the incident beam, $\theta_z=\pi$ ($\mu=-1$) to emission in the backward direction, and $\theta_z=\pi/2$ ($\mu=0$) to emission perpendicular to the beam axis. Assuming azimuthal symmetry about the beam axis, which holds by construction for the
mono-directional broomstick source, the differential solid angle subtended by a polar bin $[\theta_{i-1}, \theta_i]$ is
\begin{equation}
\Delta\Omega_i = 2\pi\left(\cos\theta_{i-1} - \cos\theta_i\right) = 2\pi\left(\mu_{i-1} - \mu_i\right),
\label{eq:domega}
\end{equation}
where $\mu = \cos\theta_z$. Because $\Delta\Omega_i$ depends on the bin location (i.e., polar bins near $\mu = \pm 1$ subtend a smaller solid angle than bins near $\mu = 0$ for a uniform grid in $\theta$) the raw bin-wise neutron counts cannot be compared directly across codes unless this geometric weighting is removed. The angular current density per unit solid angle in bin $i$ was therefore obtained as
\begin{equation}
\left.\frac{\text{dY}_n}{\text{d}\Omega}\right|_i = \frac{\text{Y}_i}{\Delta\Omega_i} = \frac{\text{Y}_i}{2\pi\left(\mu_{i-1}-\mu_i\right)},
\label{eq:dydomega}
\end{equation}
with units of neutrons per source photon per steradian (n\,$\upgamma^{-1}$\,sr$^{-1}$).

The angular tallying scheme differed among the three Monte Carlo codes and required this explicit normalization to allow a direct comparison. MCNPX and FLUKA natively score the angular current in 90 uniform bins in $\mu \in [-1, 1]$, for which $\Delta\Omega_i$ is constant and Eq.~\eqref{eq:dydomega} reduces to a uniform scaling of the raw bin counts.
OpenMC, through its \texttt{PolarFilter} tally, instead scores the current in 90 uniform bins in the polar angle $\theta_z \in [0, \pi]$; because $\Delta\Omega_i$ is not constant for a uniform grid in $\theta_z$, the OpenMC bin counts were converted to the corresponding $\mu$-bin edges and normalized using Eq.~\eqref{eq:dydomega} prior to comparison with the other codes. All angular results reported in Section~\ref{sec:results} are expressed as
$dY_n/d\Omega$ following this common normalization. A consequence of the uniform $\theta_z$ grid is that the OpenMC bins adjacent to $\mu$ = $\pm$1 subtend a much smaller solid angle than the uniform µ bins of MCNPX and FLUKA, and therefore collect fewer scores. The isolated spikes or drops visible in the outermost OpenMC bins in Fig.~\ref{fig:angular_current} are attributable to this larger statistical uncertainty and should not be interpreted as physical features.

Because the energy-resolved and angular tallies are two independent scorings of the same underlying outward neutron current, the integrated yield $\text{Y}_{tot}$ can equivalently be obtained either by summing the bin-integrated energy-resolved contributions or by integrating the angular current density of Eq.~\eqref{eq:dydomega} over the full solid angle,
\begin{equation}
\text{Y}_{tot} = \sum_{i=1}^{\text{N}_E} \text{Y}_i = \int \frac{\text{dY}_n}{\text{d}\Omega}\, \text{d}\Omega = 2\pi \int_{-1}^{1} \frac{\text{dY}_n}{\text{d}\Omega}\, \text{d}\mu.
\label{eq:ytot_dual}
\end{equation}
where $\text{N}_E$ is the number of spectral channels and Y$_i$ is the corresponding bin-integrated yield. Both estimates were computed for every numerical benchmark and calculation case as a consistency check on the reported tallies; the two agreed to within Monte Carlo statistical uncertainty in all cases. Unless otherwise noted, the integrated yields reported in Section~\ref{sec:results} correspond to the solid-angle integral of Eq.~\eqref{eq:ytot_dual}.

OpenMC and MCNPX uncertainties correspond to the standard errors reported by their respective Monte Carlo tally estimators. FLUKA boundary-crossing scores were combined across the ten independent production runs using the standard FLUKA post-processing utilities. The statistical uncertainty of $\text{Y}_{tot}$, $\sigma_{\text{Y}}$, was obtained by combining the standard errors of the angular bins in quadrature. To assess whether the difference between the OpenMC and MCNPX integrated yields is statistically significant, the difference was expressed in units of the combined standard deviation of the two calculations,
\begin{equation}
    n_\sigma = \frac{\left|\text{Y}_{tot}^{\mathrm{MCNPX}} - \text{Y}_{tot}^{\mathrm{OpenMC}}\right|}
    {\sqrt{(\sigma_{\text{Y}}^{\mathrm{MCNPX}})^2 + (\sigma_{\text{Y}}^{\mathrm{OpenMC}})^2}},
    \label{eq:nsigma}
\end{equation}
since the two calculations are statistically independent. Differences with $n_\sigma \leq 2$ are regarded as consistent with statistical fluctuations at approximately the 95\% confidence level, whereas $n_\sigma > 3$ indicates a statistically significant difference.

\subsection{First-collision analytical solution of the photoneutron yield}
\label{sec:analytical}

A first-collision analytical solution of the integrated photoneutron yield was introduced as a transport-code-independent check of the Monte Carlo calculations. The formulation considers the competition among the possible first photon interactions in the target and weights each neutron-producing photonuclear channel by its neutron multiplicity. It does not model the energy or angular distributions of the emitted neutrons and therefore provides an integral-yield check rather than a prediction of the secondary-neutron spectrum.\\
For a target composed of nuclides $k$ with atomic number densities $N_k$, the macroscopic neutron-production cross section is defined as

\begin{equation}
    \Sigma_{n}(E_\gamma)
    =
    \sum_{k} N_k
    \sum_{j}
    \nu_{jk}(E_\gamma)\,
    \sigma_{jk}^{\mathrm{pn}}(E_\gamma),
    \label{eq:macro_neutron_production}
\end{equation}

\noindent where $\sigma_{jk}^{\mathrm{pn}}$ is the microscopic cross section of the neutron-producing photonuclear reaction $j$ on nuclide $k$, and $\nu_{jk}$ is the corresponding mean neutron multiplicity. Thus, $\nu=1$ for a single-neutron-emission channel, $\nu=2$ for $(\upgamma,\text{2n})$, and $\nu=\nu_{f}(E_\gamma)$ for photofission. Microscopic cross sections for the nuclides considered in the present work are depicted in Figure~\ref{fig:xs}.

The total macroscopic photon interaction cross section entering the
first-collision balance is
\begin{equation}
    \Sigma_{\mathrm{tot}}(E_\gamma)
    =
    \sum_{k} N_k
    \left[
        \sum_{i}
        \sigma_{ik}^{\mathrm{pa}}(E_\gamma)
        +
        \sum_{r}
        \sigma_{rk}^{\mathrm{pn}}(E_\gamma)
    \right],
    \label{eq:macro_total}
\end{equation}
\noindent where $\sigma_{ik}^{\mathrm{pa}}$ denotes the microscopic photoatomic cross sections, including coherent and incoherent scattering, photoelectric absorption, and pair production, while the second sum contains the photonuclear channels included in the evaluated dataset.

The mean number of neutrons generated per first photon interaction is
\begin{equation}
    w(E_\gamma)
    =
    \frac{\Sigma_{n}(E_\gamma)}
         {\Sigma_{\mathrm{tot}}(E_\gamma)}.
    \label{eq:w}
\end{equation}

Unlike a conventional reaction probability, $w(E_\gamma)$ includes the neutron multiplicity and therefore represents the expected number of neutrons produced per first photon interaction. Figure~\ref{fig:w} shows the different $w$ factors for the nuclides considered in the present work. It is interesting to note that for the single-nuclide targets considered in this study, the atomic density appears in both the numerator and denominator of Eq.~\eqref{eq:w} and consequently cancels exactly.

For a target of finite axial length $L$, the corresponding yield per
incident photon is
\begin{equation}
    \text{Y}_n(L)
    =
    \int_{E_{\min}}^{E_{\max}}
    \frac{\text{d}N_\gamma}{\text{d}E_\gamma}\,
    \left[
        1-
        \exp\left(
            -\Sigma_{\mathrm{tot}}(E_\gamma)L
        \right)
    \right]
    w(E_\gamma)\,
    \text{d}E_\gamma,
    \label{eq:yield_finite_length}
\end{equation}
\noindent where $\frac{\text{d}N_\gamma}{\text{d}E_\gamma}$ is the normalized incident photon energy distribution (also shown in Figure~\ref{fig:w}):
\begin{equation}
    \int_{E_{\min}}^{E_{\max}}
    \frac{\text{d}N_\gamma}{\text{d}E_\gamma}\,\text{d}E_\gamma
    =
    1.
    \label{eq:source_normalization}
\end{equation}

The broomstick targets are extremely thin in the radial direction but long in the beam direction. They are designed so that a primary photon can undergo a first interaction along the target axis, whereas the particles generated or scattered in that interaction leave the target almost immediately through the lateral surface. In the axially thick limit:
\begin{equation}
    1-
    \exp\left[
        -\Sigma_{\mathrm{tot}}(E_\gamma)L
    \right]
    \simeq
    1,
    \label{eq:thick_limit}
\end{equation}
and Eq.~\eqref{eq:yield_finite_length} reduces to
\begin{equation}
    \text{Y}_n
    \simeq
    \int_{E_{\min}}^{E_{\max}}
        \frac{\text{d}N_\gamma}{\text{d}E_\gamma}\,
    w(E_\gamma)\,
    \text{d}E_\gamma.
    \label{eq:Yn}
\end{equation}

Equation~\eqref{eq:yield_finite_length} represents the general first-collision expression for a target of finite axial length. When the photon interaction probability over the target length is numerically indistinguishable from unity, Eq.~\eqref{eq:yield_finite_length} reduces to the axially thick approximation of Eq.~\eqref{eq:Yn}. The applicability of this approximation must therefore be established by evaluating the finite-length interaction factor over the photon-energy range contributing to each case. 
In the idealized single-collision limit of the broomstick geometry, neutrons produced in the first interaction escape through the infinitesimal lateral dimension of the target with negligible secondary transport. The calculated first-collision production yield can consequently be compared with the integrated outward neutron current obtained from the Monte Carlo calculations.

The photonuclear cross sections used in the analytical solution were taken from the ENDF7u evaluation adopted in the common-data OpenMC--MCNPX comparison. Consequently, the analytical result provides an additional check of the evaluated-data calculations, but it is not expected to reproduce the native FLUKA photonuclear model channel by channel, nor the OpenMC yield with the IAEA/PD-2019 nuclear data. As a first approximation, no covariance-based nuclear-data uncertainty was propagated through the calculation. Therefore, the comparison concerns the central values of the adopted evaluations.

\subsection{Photon source spectra}
\label{sec:sources}

Two classes of incident photon sources were considered: monoenergetic beams and a continuous, LINAC-representative spectrum. In all cases, the photons originated at $z=0$ and propagated parallel to the positive target axis.

\subsubsection{Monoenergetic sources}

For the monoenergetic cases, the normalized source distribution is approximated by a Dirac delta function
\begin{equation}
    \frac{\text{d}N_\gamma}{\text{d}E_\gamma}
    \propto
    \delta(E_\gamma -E_0),
    \label{eq:mono}
\end{equation}
\noindent where $E_0=5$~MeV for the $^{2}$H and $^{9}$Be targets and $E_0=15$~MeV for $^{238}$U. Substitution into Eq.~\eqref{eq:yield_finite_length} gives
\begin{equation}
    \text{Y}_n(E_0,L)
    =
    \left[
        1-
        \exp\left(
            -\Sigma_{\mathrm{tot}}(E_0)L
        \right)
    \right]
    w(E_0).
    \label{eq:mono_yield}
\end{equation}

In the axially thick limit, $\Sigma_{\mathrm{tot}}(E_0)L\gg1$, this expression reduces to
\begin{equation}
    \text{Y}_n(E_0,L)
    \simeq
    w(E_0).
    \label{eq:mono_yield_thick}
\end{equation}

\subsubsection{Continuous LINAC-representative source}

The continuous source was based on the simple Kramers bremsstrahlung approximation~\cite{Kramers1923},
\begin{equation}
    \frac{\text{d}N_\gamma}{\text{d}E_\gamma}
    \propto
    \frac{E_{\mathrm{end}}-E_\gamma}{E_\gamma},
    \qquad
    E_{\min}\le E_\gamma\le E_{\mathrm{end}},
    \label{eq:kramers_unnormalized}
\end{equation}
\noindent with
\begin{equation}
    E_{\min}=1~\mathrm{MeV},
    \qquad
    E_{\mathrm{end}}=20~\mathrm{MeV}.
\end{equation}

and in this case the corresponding Y$_n$ value is evaluated through numerical integration of Equation~\eqref{eq:Yn}.

\begin{figure}[htbp]
    \centering
    \includegraphics[width=0.90\linewidth]{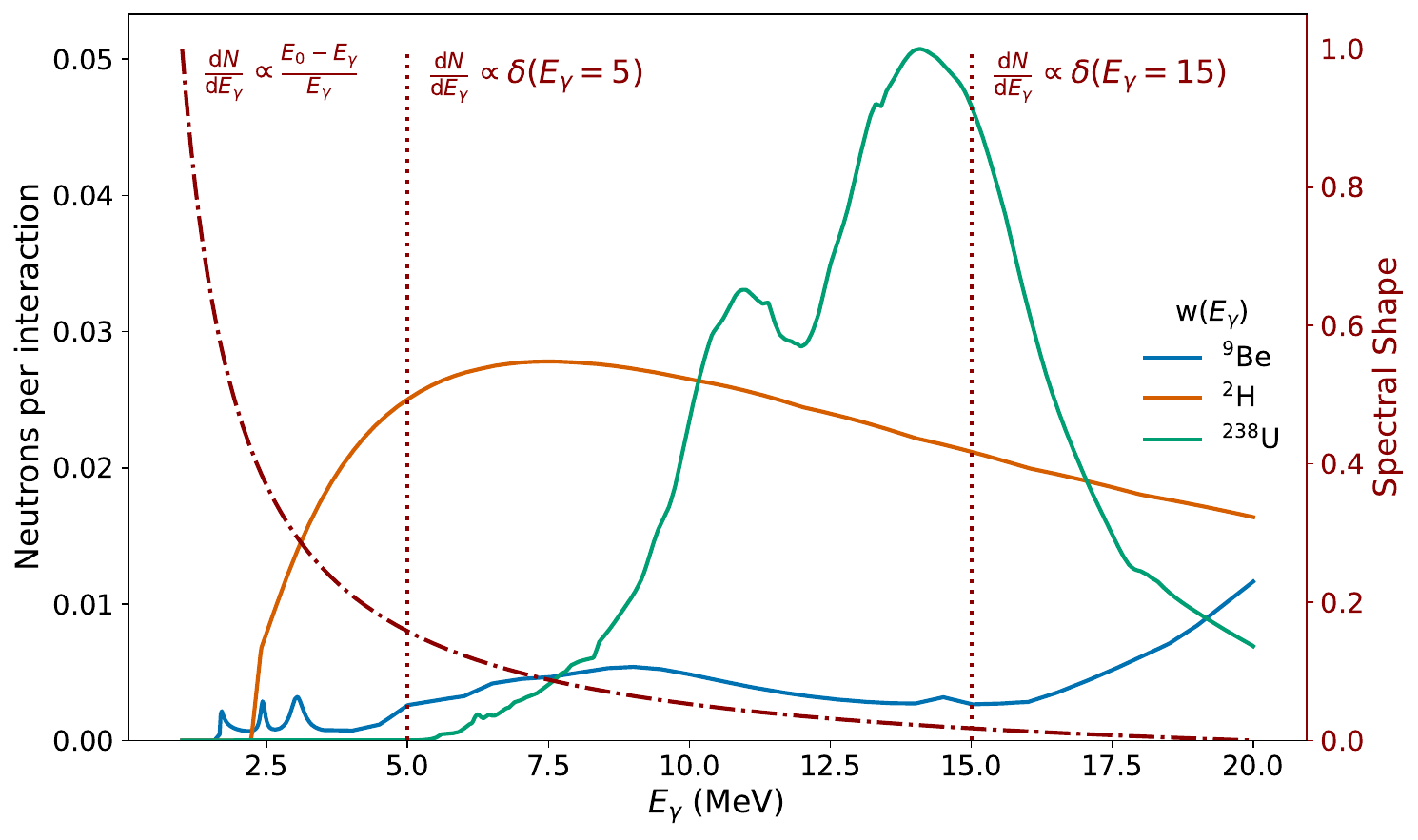}
    \caption{Expected number of neutrons produced per first photon
    interaction, $w(E_\gamma)$, for $^{2}$H, $^{9}$Be, and
    $^{238}$U using the ENDF7u cross section data. The curves are shown together with the normalized
    LINAC-representative Kramers spectrum ($E_{\mathrm{end}}=20$~MeV) and the monoenergetic source energies of 5 and 15~MeV.}
    \label{fig:w}
\end{figure}

Figure~\ref{fig:w} shows $w(E_\gamma)$ for the three target materials, together with the normalized continuous source and the two monoenergetic photon energies. The figure illustrates the different
regions of the photonuclear response sampled by the two source classes.

\section{Results}
\label{sec:results}

\subsection{Energy-resolved neutron yields}
\label{sec:spectral_results}

\begin{figure}[htbp]
    \centering
    \includegraphics[width=\linewidth]
    {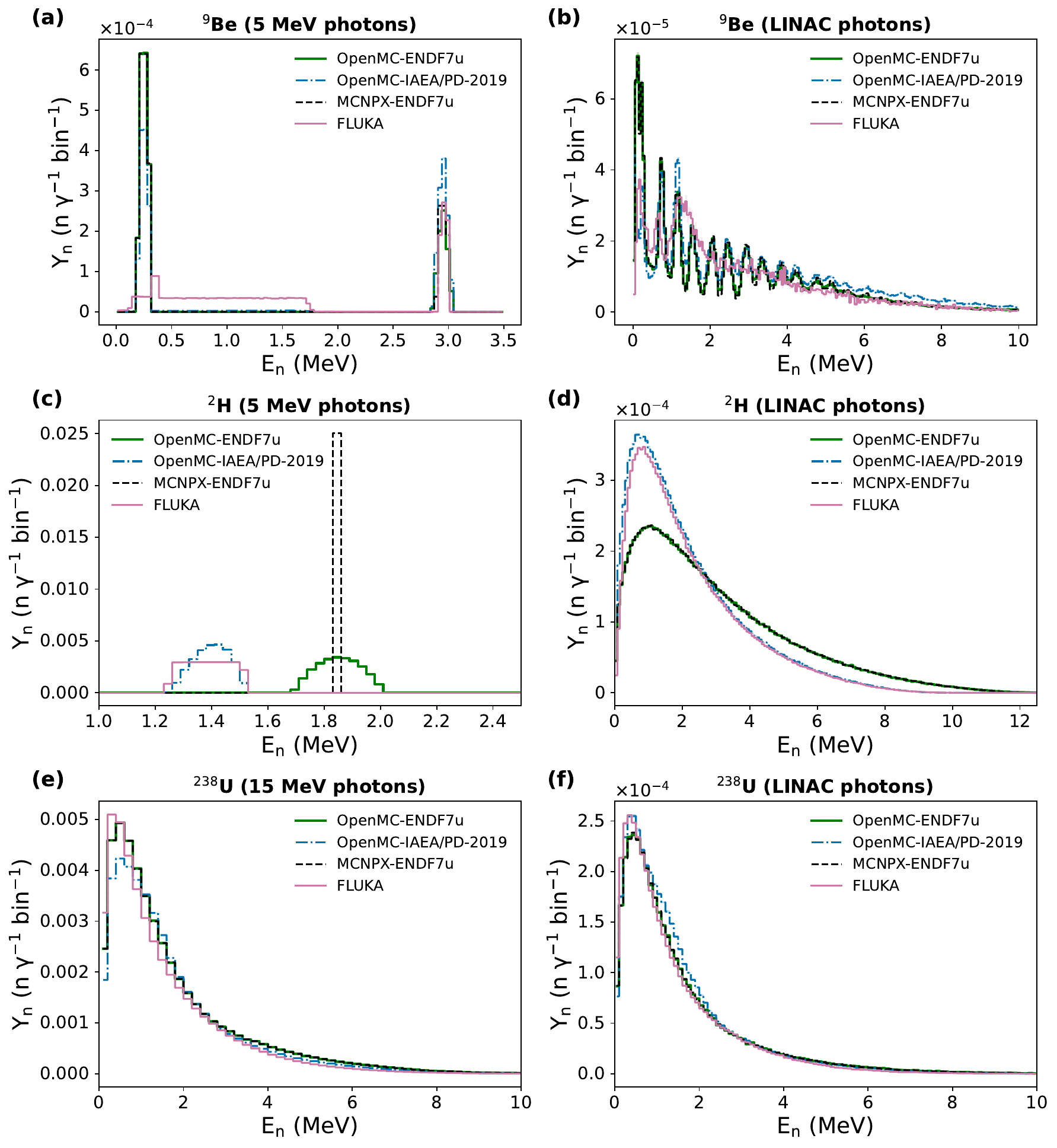}
    \caption{Energy-resolved neutron yields for the six broomstick cases: (a) $^{9}$Be irradiated by 5~MeV photons; (b) $^{9}$Be irradiated by the continuous LINAC-representative source; (c) $^{2}$H irradiated by 5~MeV photons; (d) $^{2}$H irradiated by the continuous source; (e) $^{238}$U irradiated by 15~MeV photons; and (f) $^{238}$U irradiated by the continuous source. OpenMC and MCNPX calculations labeled ENDF7u use the common evaluated photonuclear data. OpenMC with IAEA/PD-2019 illustrates the nuclear-data sensitivity, whereas FLUKA uses its native photonuclear interaction treatment.}
    \label{fig:all_spectra}
\end{figure}

Figure~\ref{fig:all_spectra} compares the energy-binned neutron yields predicted for all six configurations, for both the monoenergetic and the continuous LINAC-representative sources. Each plotted value represents the neutron yield integrated within the corresponding energy bin and normalized per source photon. All energy-resolved yields are reported with their 1$\sigma$ statistical uncertainty, shown as shaded bands. With $10^{8}$ source photons, the relative uncertainty is below 0.5\% in all energy bins, so the bands are generally narrower than the line width and the differences discussed below are well above the statistical uncertainty.

For $^{9}$Be at 5~MeV in Fig.~\ref{fig:all_spectra}(a), the OpenMC--ENDF7u and MCNPX spectra reproduce the same two dominant neutron-energy groups, located near 0.25 and 2.9~MeV. These structures arise from the restricted kinematics of the $^{9}$Be photonuclear breakup channels at the selected incident energy. The relative magnitudes of the two groups change when IAEA/PD-2019 data are used in OpenMC. FLUKA predicts a substantially smoother low-energy contribution, extending over a broader interval. The $^{9}$Be case therefore shows that similar integrated production can coexist with appreciable differences in the sampled final-state spectrum. Folding the photonuclear response over the continuous source broadens the neutron distributions and reduces several of the localized differences observed under monoenergetic irradiation: in Fig.~\ref{fig:all_spectra}(b), OpenMC--ENDF7u and MCNPX closely reproduce the sequence of structures generated by the convolution of the source distribution with the energy-dependent photonuclear response, the FLUKA spectrum is smoother and decreases more rapidly at high neutron energies, and OpenMC with IAEA/PD-2019 predicts a larger high-energy contribution.\\

The 5-MeV $^{2}$H numerical benchmark shown in Fig.~\ref{fig:all_spectra}(c) exhibits the largest spectral differences among the six cases. From two-body kinematics, the neutrons emitted at 5~MeV should lie between approximately 1.25 and 1.52~MeV, with a mean of about 1.4~MeV, as reproduced by OpenMC with IAEA/PD-2019 and by FLUKA. Both ENDF7u calculations instead place the neutrons near 1.85~MeV. This shift originates in the ENDF7u data rather than in the transport codes: NJOY versions prior to NJOY2016.66 computed the outgoing energy of two-body photonuclear reactions with the expression valid for an incident neutron, which for $^{2}$H overestimates the neutron energy (1.85 instead of 1.37~MeV in the center-of-mass frame)~\cite{NJOY2016_releasenotes}. The error is negligible for heavier nuclides and does not affect the reaction cross section, and hence the integrated yield. Libraries processed with later NJOY versions, such as IAEA/PD-2019, are not affected. The different line shape of the two ENDF7u results has a separate origin. The tabulated energy refers to the center-of-mass frame: OpenMC converts it to the laboratory frame accounting for the photon momentum, which spreads the neutron energies over approximately 1.70--2.01~MeV, whereas MCNPX emits all neutrons at the center-of-mass energy, producing a single narrow line. A similar behavior has recently been reported for MCNP6~\cite{Garnaud2026}. As shown below, these spectral differences are considerably larger than the corresponding differences in the integrated yield. Under the continuous source (Fig.~\ref{fig:all_spectra}(d)), OpenMC--ENDF7u and MCNPX are nearly superimposed over the full energy range, while OpenMC with IAEA/PD-2019 and FLUKA predict a larger low-energy maximum, followed by a more rapid decrease. The harder ENDF7u spectra are consistent with the same processing issue, whereas the line-shape difference observed at a single incident energy is averaged out by integration over the continuous photon spectrum.\\

For $^{238}$U irradiated at 15~MeV (Fig.~\ref{fig:all_spectra}(e)), all calculations predict a broad neutron distribution dominated by low-energy emission and followed by a decreasing high-energy tail. The overall shapes are more similar than in the light-nuclide cases, although differences remain in the magnitude of the low-energy peak and in the tail. Because $(\upgamma,\text{n})$, $(\upgamma,\text{2n})$, and photofission contribute simultaneously, agreement in the total spectrum does not necessarily imply agreement in the individual reaction channels. The four $^{238}$U continuous-source spectra show comparatively close agreement in Fig.~\ref{fig:all_spectra}(f); differences are primarily concentrated below approximately 2~MeV and become small in the decreasing high-energy tail.

\subsection{Angular distribution of the emitted neutron current}
\label{sec:angular_results}

\begin{figure}[htbp]
    \centering
    \includegraphics[width=\textwidth]{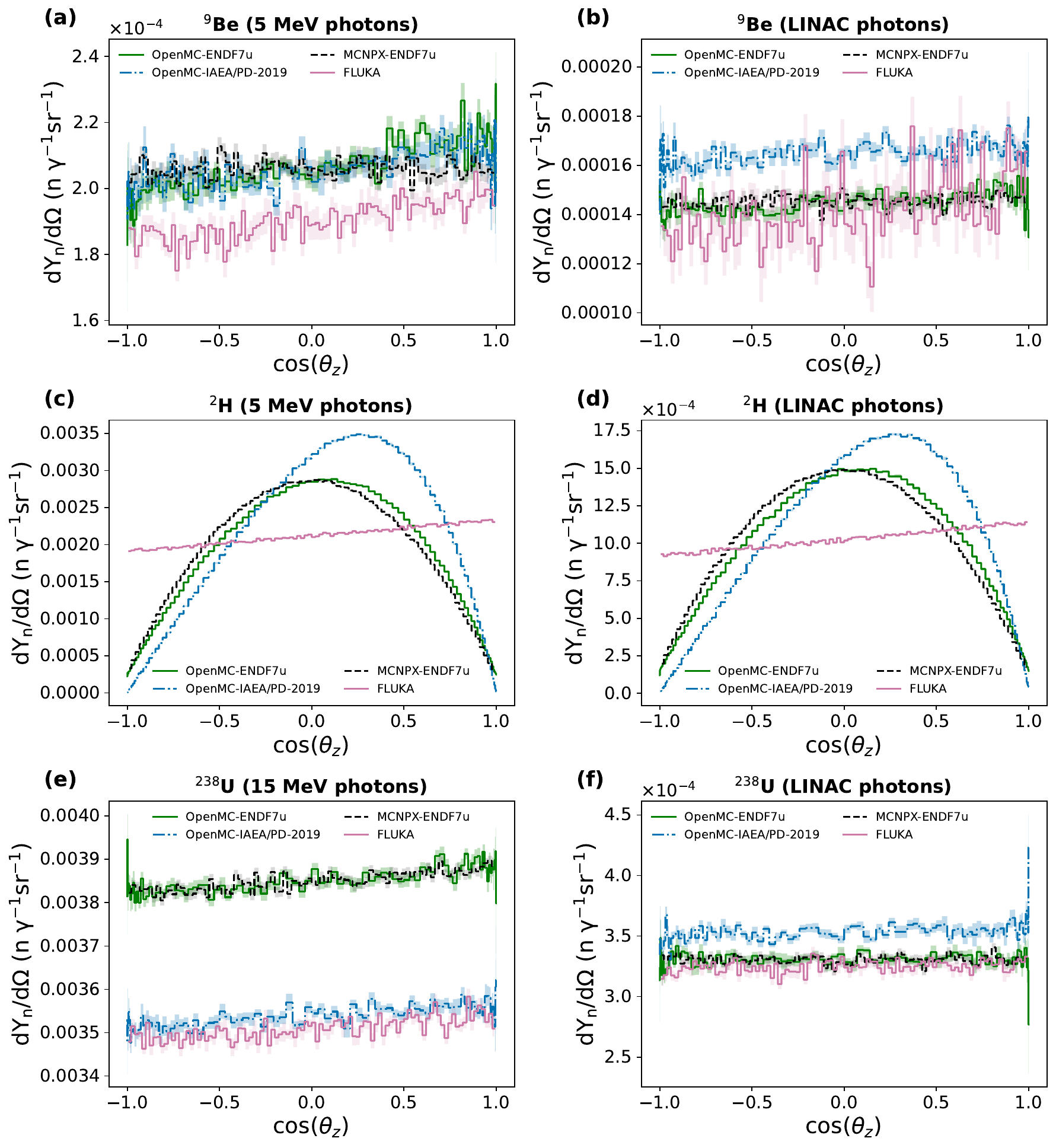}
    \caption{Angular distribution of the escaping neutron current, $\text{dY}_n/\text{d}\Omega$, as a
    function of $\mu = \cos\theta_z$, for the six broomstick benchmarks: (a) $^9$Be
    irradiated by 5~MeV photons; (b) $^9$Be irradiated by the continuous LINAC-representative
    source; (c) $^2$H irradiated by 5~MeV photons; (d) $^2$H irradiated by the continuous
    source; (e) $^{238}$U irradiated by 15~MeV photons; and (f) $^{238}$U irradiated by the
    continuous source. OpenMC and MCNPX calculations labeled ENDF7u use the common evaluated
    photonuclear data. OpenMC with IAEA/PD-2019 illustrates the nuclear-data sensitivity,
    whereas FLUKA uses its native photonuclear interaction treatment.}
    \label{fig:angular_current}
\end{figure}

Figure~\ref{fig:angular_current} reports the angular current density, $\text{dY}_n/\text{d}\Omega$ defined in Eq.~\eqref{eq:dydomega}, as a function of $\mu = \cos\theta_z$ for all six numerical benchmark configurations and all four calculation cases (OpenMC-ENDF7u, OpenMC-IAEA/PD-2019, MCNPX-ENDF7u, and FLUKA). As for the energy spectra, the angular distributions are reported with their 1$\sigma$ statistical uncertainty, which is below 1\% in all significant angular bins.

For $^9$Be, the ENDF7u $(\gamma,n2\alpha)$ neutrons are isotropic in the center-of-mass frame. OpenMC transforms them to the laboratory frame, which produces a forward-backward asymmetry of about $\pm$4\% because of the low energy of most emitted neutrons, whereas MCNPX emits them isotropically in the laboratory frame, as for $^2$H. OpenMC-ENDF7u and MCNPX-ENDF7u agree in magnitude, but their shapes differ slightly: the MCNPX current is flat within statistical fluctuations, whereas the OpenMC current increases mildly toward forward angles. OpenMC-IAEA/PD-2019 follows the same trend as OpenMC-ENDF7u, while FLUKA predicts a systematically lower current (by about 8\%, consistent with its integrated yield) with a similar mild forward increase. Under the continuous LINAC source (Fig.~\ref{fig:angular_current}(b)), the currents remain nearly isotropic, with the OpenMC calculations again increasing slightly toward forward angles; OpenMC-ENDF7u and MCNPX-ENDF7u agree in magnitude, OpenMC-IAEA/PD-2019 is shifted to a systematically higher magnitude, and the FLUKA distribution, while statistically noisier due to the smaller number of scored events per angular bin, is centered on a comparable mean value.
The $^2$H numerical benchmarks (Fig.~\ref{fig:angular_current}(c,d)) exhibit the most pronounced angular-shape differences among the six cases. All ENDF7u- and IAEA/PD-2019-based OpenMC and MCNPX calculations predict a strongly side-peaked distribution, consistent with the two-body kinematics of the $^2$H($\upgamma$,n) reaction. 

OpenMC-ENDF7u and MCNPX-ENDF7u predict the same peak height and width at both source energies, confirming that the same perpendicular-peaked angular law is sampled by both codes when the same evaluated data are used. The OpenMC distribution is, however, slightly shifted toward forward angles: relative to MCNPX, which is symmetric about $\mu$ = 0, it is about 12\% lower at $\mu$ = -0.75 and about 11\% higher at $\mu$ = +0.75 for the 5-MeV source. For both light nuclides, this forward shift is consistent with the center-of-mass-to-laboratory transformation that OpenMC applies and MCNPX does not, which also spreads the OpenMC neutron energies in Fig.~\ref{fig:all_spectra}(c) (Section~\ref{sec:spectral_results}).

OpenMC-IAEA/PD-2019 preserves the side-peaked character but predicts a higher and narrower peak than the ENDF7u-based calculations. FLUKA, in contrast, predicts a markedly different angular shape: a weakly forward anisotropy distribution that lies below the ENDF7u-based peak value. This indicates that the native FLUKA photonuclear treatment of the $^2$H two-body breakup does not reproduce the strongly side-peaked kinematics captured by the ENDF7u and IAEA/PD-2019 evaluated angular distributions, despite the comparatively modest differences observed in the integrated yield for this case (Section~\ref{sec:integrated_results}).\\

For $^{238}$U at 15 MeV (Fig.~\ref{fig:angular_current}(e)), all four calculations show a nearly isotropic distribution with a weak increase toward forward angles, of about 1\% between the backward and forward hemispheres. OpenMC-ENDF7u and MCNPX-ENDF7u are superimposed across the full angular range, whereas OpenMC-IAEA/PD-2019 and FLUKA lie close to each other at a systematically lower magnitude, consistent with their lower integrated yields (Section~\ref{sec:integrated_results}). Under the continuous LINAC source (Fig.~\ref{fig:angular_current}(f)), the distributions become flat. OpenMC-ENDF7u and MCNPX-ENDF7u are superimposed, FLUKA lies about 2\% lower, and OpenMC-IAEA/PD-2019 lies systematically higher, consistent with the integrated yields (Section~\ref{sec:integrated_results}).

\subsection{Integrated neutron-yield comparison}
\label{sec:integrated_results}

\begin{figure}[htbp]
    \centering
    \includegraphics[width=\linewidth]
    {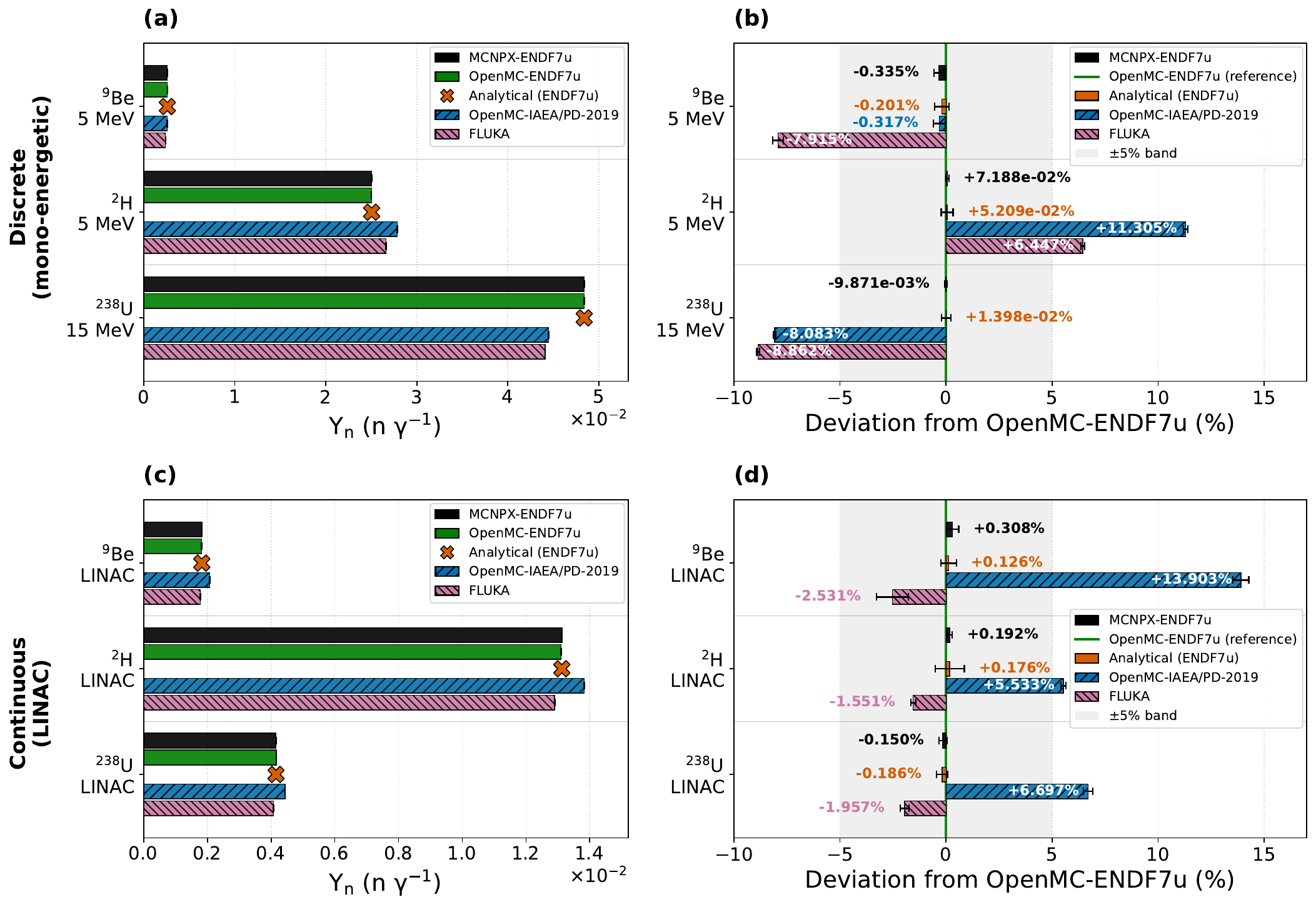}
    \caption{Integrated neutron-yield comparison for (a,b) the three monoenergetic numerical cases and (c,d) the three continuous-source cases. Panels (a) and (c) show the absolute yields per source photon, with the first-collision analytical solution marked by a cross; panels (b) and (d) report the relative deviation from OpenMC–ENDF7u, where the analytical solution is shown as a solid bar. Bars follow the same vertical order in all panels. Solid bars identify calculations based on the common ENDF7u evaluation, including the analytical solution; hatched bars identify either the IAEA/PD-2019 evaluation or the native FLUKA photonuclear treatment. The shaded region represents a $\pm$5\% interval about OpenMC–ENDF7u.}
    \label{fig:integrated_yields}
\end{figure}

Figure~\ref{fig:integrated_yields} reports the integrated neutron yields and their deviations from the OpenMC-ENDF7u calculation, used as reference. The shaded region denotes a deviation of $\pm5\%$ from the reference. As described in Section~\ref{sec:mc_implementations}, Y$_{tot}$ is obtained by integrating the angular current density over the full solid angle (Eq.~\eqref{eq:ytot_dual}).

The most direct verification of the new OpenMC photonuclear branch is provided by comparison with MCNPX under the same ENDF7u photonuclear evaluation. Table~\ref{tab:ytot_openmc_mcnpx} reports the integrated yields calculated by the two codes, the relative deviation of MCNPX from OpenMC, $\Delta$, and the number of combined standard deviations, $n_\sigma$, defined in Eq.~\eqref{eq:nsigma}. The two codes agree to within approximately 0.35\% in all six numerical benchmark configurations; the largest deviations occur for the two $^{9}$Be cases ($-0.34\%$ monoenergetic, $+0.31\%$ continuous), while the $^{2}$H and $^{238}$U cases differ by less than 0.2\%. In all cases $n_\sigma \leq$ 2, so the residual differences are consistent with the statistical uncertainty of the two calculations and do not indicate a systematic difference between the two implementations. For monoenergetic irradiation, the first-collision analytical solution differs from OpenMC–ENDF7u by approximately -0.20\% for $^9$Be, +0.05\% for $^2$H, and +0.01\% for $^{238}$U; for the continuous source, the corresponding deviations are approximately +0.13\%, +0.18\%, and -0.19\%. This agreement to within about 0.2\% in all six cases confirms that the integrated neutron-production yield is consistently reproduced, despite the spectral differences observed for $^2$H (see Section~\ref{sec:spectral_results}).

\begin{table}[htbp]
    \centering
    \small
    \caption{Integrated neutron yields per source photon, $\text{Y}_{tot}$, calculated by OpenMC and MCNPX with the common ENDF7u photonuclear evaluation ($1\sigma$ statistical uncertainties). $\Delta = 100\,(\text{Y}_{tot}^{\mathrm{MCNPX}}/\text{Y}_{tot}^{\mathrm{OpenMC}} - 1)$ is the relative deviation of MCNPX from OpenMC, and $n_\sigma$ is the difference in units of the combined standard deviation (Eq.~\eqref{eq:nsigma}).}
    \label{tab:ytot_openmc_mcnpx}
    \begin{tabular}{llcccc}
        \toprule
        \textbf{Nuclide} & \textbf{Source} & \textbf{OpenMC-ENDF7u} & \textbf{MCNPX-ENDF7u} & \textbf{$\Delta$} & \textbf{$n_\sigma$}\\
         & & \textbf{(n\,$\upgamma^{-1}$)} & \textbf{(n\,$\upgamma^{-1}$)} & \textbf{(\%)} & \\
        \midrule
        $^{9}$Be & 5 MeV & $(2.5961 \pm 0.0049) \times 10^{-3}$ & $(2.5874 \pm 0.0036) \times 10^{-3}$ & $-0.34$ & 1.43 \\
        $^{2}$H & 5 MeV & $(2.5022 \pm 0.0015) \times 10^{-2}$ & $(2.5040 \pm 0.0011) \times 10^{-2}$ & $+0.07$ & 0.96 \\
        $^{238}$U & 15 MeV & $(4.8384 \pm 0.0022) \times 10^{-2}$ & $(4.8379 \pm 0.0016) \times 10^{-2}$ & $-0.01$ & 0.18 \\
        \midrule
        $^{9}$Be & LINAC & $(1.8201 \pm 0.0043) \times 10^{-3}$ & $(1.8257 \pm 0.0030) \times 10^{-3}$ & $+0.31$ & 1.07 \\
        $^{2}$H & LINAC & $(1.3103 \pm 0.0012) \times 10^{-2}$ & $(1.31279 \pm 0.00081) \times 10^{-2}$ & $+0.19$ & 1.78 \\
        $^{238}$U & LINAC & $(4.1604 \pm 0.0064) \times 10^{-3}$ & $(4.1541 \pm 0.0046) \times 10^{-3}$ & $-0.15$ & 0.80 \\
        \bottomrule
    \end{tabular}
\end{table}

\subsubsection{Sensitivity to the photonuclear evaluation}

Changing the OpenMC photonuclear library from ENDF7u to IAEA/PD-2019 produces differences that are generally larger than the OpenMC--MCNPX code-to-code differences obtained under common data.
For the monoenergetic cases, the IAEA/PD-2019 results differ from OpenMC–ENDF7u by approximately -0.32\% for $^9$Be, +11.3\% for $^2$H, and -8.1\% for $^{238}$U. For the continuous source, the corresponding deviations are approximately +13.9\%, +5.5\%, and +6.7\%. The sensitivity depends strongly on both the nuclide and the source spectrum, reflecting differences between the evaluations over the energy interval sampled in each case.
The library sensitivity is particularly evident for $^2$H under monoenergetic irradiation and for $^9$Be under continuous irradiation, and remains significant for $^{238}$U under both sources.

\subsubsection{Independent FLUKA comparison}
For the monoenergetic cases, the FLUKA yields differ from OpenMC–ENDF7u by approximately -7.9\% for $^9$Be, +6.4\% for $^2$H, and -8.9\% for $^{238}$U. The discrepancies decrease after folding over the continuous photon source, where FLUKA differs from OpenMC–ENDF7u by approximately -2.5\% for $^9$Be, -1.6\% for $^2$H, and -2.0\% for $^{238}$U.\\
The improved integral agreement under continuous irradiation indicates that positive and negative energy-dependent differences are partially averaged over the incident spectrum. It does not imply equivalence of the underlying photonuclear cross sections, final-state models, or reaction-channel contributions.

\section{Discussion}
\label{sec:discussion}

The close agreement between OpenMC and MCNPX under common ENDF7u data, statistically consistent in all six configurations (Table~\ref{tab:ytot_openmc_mcnpx}), provides direct evidence that the unofficial OpenMC photonuclear branch (version 0.16.1-dev138) reproduces the integrated neutron-production response of an established reference code, across three nuclides and two source classes. The energy-resolved and angular results refine this picture. 
For $^2$H at 5 MeV, OpenMC and MCNPX predict the same total neutron production but visibly different spectral line shapes. Their angular currents, sharply peaked near $\theta_z \approx 90^\circ$ by the two-body kinematics of $^2$H($\upgamma$,n), have the same peak height and width, which confirms correct sampling of the angular law encoded in the evaluated data, not merely the correct event count; this is a stronger verification statement than the sub-percent agreement in Y$_{tot}$ alone. The remaining differences, a broader energy line and a slight forward shift of the angular current in OpenMC, are both consistent with the center-of-mass-to-laboratory transformation, which OpenMC applies and MCNPX does not. Because the $^{2}$H($\gamma$,n) angular law (MT=50) is represented as an \textit{uncorrelated angle-energy distribution} in both ENDF7u and IAEA/PD-2019, the peak differences between the two evaluations reflect genuine differences in the tabulated angular values rather than a change in representation format. FLUKA, using its native model, instead predicts a markedly weaker, forward-shifted anisotropy despite an integrated yield within about 6\% of the OpenMC-ENDF7u calculations: here the angular shape, not the total yield, is the more discriminating observable, since a shielding or dosimetry calculation relying on FLUKA would place a comparable number of neutrons in the wrong directions rather than the wrong number overall.

 A similar pattern emerges for $^9$Be. FLUKA predicts a smoother, less structured energy spectrum than the evaluated-data calculations and a lower angular current, but with a similar, nearly isotropic angular shape, so that its differences for $^9$Be lie in the energy spectrum and in the magnitude rather than in the angular emission. Changing the representation of the dominant MT=29 channel between libraries (uncorrelated in ENDF7u, correlated in IAEA/PD-2019) leaves the angular shape nearly unchanged. Under common ENDF7u data, OpenMC and MCNPX agree in integrated yield and energy spectrum, while the OpenMC angular current increases slightly toward forward angles; as for $^2$H, this is consistent with the center-of-mass-to-laboratory transformation applied only by OpenMC. For $^{238}$U, OpenMC and MCNPX agree in integrated yield, energy spectrum, and angular distribution under common ENDF7u data, even though the response combines several channels with distinct representations: the dominant ($\upgamma$,n)/($\upgamma$,2n) channels are sampled through a Kalbach-Mann correlated energy-angle law, while photofission uses a separate Maxwellian-plus-angle treatment in ENDF7u (IAEA/PD-2019 instead represents all channels through a general correlated angle-energy law). For this heavy nucleus the motion of the center of mass is negligible, and the agreement extends the verification to the multi-channel sampling logic of the unofficial branch. Taken together, these three cases confirm that integral neutron production is the most robust and most easily verified observable, while the energy and angular distributions are correspondingly more sensitive tests, able to expose implementation or data differences that an integral-only comparison would miss, consistent with earlier photonuclear benchmark studies~\cite{BarberGeorge,Tuyet2023Comparison}.

The comparison between ENDF7u and IAEA/PD-2019 shows that the underlying photonuclear evaluation can influence the calculated response more strongly than the difference between the two code implementations: replacing ENDF7u with IAEA/PD-2019 changes the integrated yield by several percent or more in several cases (most notably monoenergetic $^2$H and continuous-spectrum $^9$Be, but also $^{238}$U under both sources), whereas OpenMC and MCNPX provide consistent results, within a few tenths of a percent of each other when both use ENDF7u. A disagreement caused by different evaluated cross sections does not imply an implementation error, and conversely, agreement between codes using the same library cannot establish that the library itself is accurate relative to experimental data. The known deficiencies reported for several photonuclear evaluations, including ENDF7u, remain a distinct source of uncertainty~\cite{Sari2023,Kawano2020IAEA}. The first-collision analytical solution remains a useful integral-only check on this evaluated-data uncertainty rather than a substitute for it. Under common ENDF7u data, it agrees with OpenMC–ENDF7u to within about 0.2\% in all six cases (Section~\ref{sec:integrated_results}).

Two limitations define the scope of these conclusions. First, the broomstick geometry is deliberately idealized and suppresses secondary transport, self-shielding, and multiple interactions, making it well suited to implementation verification but not representative of a practical photoneutron target. Second, the present comparisons rely primarily on evaluated nuclear data and other transport codes and do not establish agreement with experimental measurements. Accordingly, the present work should be regarded as a focused verification of the unofficial OpenMC photonuclear branch rather than as a general validation of its predictive accuracy.

\section{Conclusions}
\label{sec:conclusions}
This work presented a focused verification of the photonuclear capability implemented in an unofficial OpenMC development branch. Six broomstick numerical benchmarks were examined using $^{2}$H, $^{9}$Be, and $^{238}$U targets irradiated by either monoenergetic photons or a continuous LINAC-representative spectrum. OpenMC was compared with MCNPX under common ENDF7u photonuclear data, with FLUKA using its native photonuclear models, and with a first-collision analytical solution for the integrated neutron yield. Additional OpenMC calculations using IAEA/PD-2019 were used to quantify sensitivity to the adopted photonuclear evaluation.

Under common ENDF7u data, the integrated neutron yields predicted by OpenMC and MCNPX agreed to within a few tenths of a percent for all six configurations, consistent with the statistical uncertainty of the calculations. This sub-percent agreement across light- and heavy-nuclide targets, different reaction mechanisms, and two source classes verifies the integrated neutron-production response of the unofficial OpenMC implementation. The energy-resolved comparisons, however, showed that agreement in the integrated yield does not necessarily imply agreement in the emitted neutron spectrum: the clearest example was the monoenergetic $^2$H case, where OpenMC and MCNPX predicted the same total neutron production with markedly different spectral widths, while IAEA/PD-2019 and FLUKA placed the emission line at a lower energy, with further differences observed in the $^9$Be breakup spectra and the low-energy region of the $^{238}$U distributions.

The angular comparison reinforced this conclusion while adding a further distinction. Under common ENDF7u data, OpenMC and MCNPX reproduced matching angular current distributions for $^{238}$U, despite its composite, multi-channel energy–angle representation, as well as the same peak height and width of the pronounced side-peaked ($\theta_z \approx 90^\circ$) emission pattern characteristic of the $^{2}$H($\gamma$,n) two-body kinematics, confirming that the correlated or uncorrelated angle-energy sampling encoded in the evaluated data is reproduced consistently by the new implementation. For the light nuclides, the OpenMC angular currents were slightly shifted toward forward angles relative to MCNPX despite sub-percent agreement in integrated yield, consistent with the center-of-mass-to-laboratory transformation that OpenMC applies and MCNPX does not. FLUKA, by contrast, predicted a markedly different angular shape for $^2$H, a nearly flat angular current with only a weak forward increase in place of the pronounced side-peaked distribution obtained with ENDF7u and IAEA/PD-2019, despite a comparable integrated yield. Angular and integrated-yield agreement should therefore be regarded as distinct verification criteria.

The OpenMC calculations using IAEA/PD-2019 demonstrated that nuclear-data differences can exceed the code-to-code differences obtained under a common evaluation, with changes relative to the OpenMC-ENDF7u reaching approximately 14\% in the cases considered; discrepancies between photonuclear calculations must therefore be separated into contributions from implementation, evaluated nuclear data, and final-state modeling. FLUKA's integrated neutron yields, relative to OpenMC–ENDF7u, differed by approximately 6–9\% for the monoenergetic cases and by about 2.5\% for the continuous-source cases.

The present results support the correctness of the integrated photonuclear neutron-production capability in the unofficial OpenMC branch under the conditions examined, within the limitations discussed in Section~\ref{sec:discussion}. Further work should extend the verification to additional nuclides and reaction thresholds, to double-differential (energy--angle) neutron distributions, and to geometries in which secondary transport is significant. Heavy non-fissile targets such as lead will be addressed in a forthcoming validation study against experimental photoneutron measurements, since experimental comparisons of integrated yields, spectra, and angular distributions are ultimately required to establish predictive accuracy. Subject to these further tests, the agreement obtained here provides a strong technical basis for continued development, independent review, and possible future integration of photonuclear physics into the official OpenMC distribution.

\pagebreak 

\appendix
\section{Comparison between MCNPX and MCNP6.2}
\label{appendixA}

Because the primary evaluated-data verification presented in this work relies on MCNPX 2.7.0, it is important to confirm that the photonuclear transport results are not specific to that particular code version. MCNPX and MCNP6 share a common lineage~\cite{goorley2012mcnp6}, and the photonuclear capability in both codes relies on the same tabulated ACE-format data
processed from evaluated photonuclear libraries, including ENDF7u~\cite{ENDFVII}. Photonuclear events are sampled using the ISPN option on the PHYS:P card in both codes, and reaction selection, secondary-neutron energy sampling, and angular sampling from the tabulated data are handled through the same underlying data-driven scheme rather than through a physics event generator~\cite{app_lanl_photonuclear}.

In particular, MCNPX 2.7.0 and MCNP6.2 read the same ACE secondary-distribution laws for photonuclear data. With ENDF7u this gives an essentially isotropic neutron emission for $^9$Be and $^{238}$U, while for $^2$H the angular distribution of the two-body ($\upgamma$,n) channel produces the side-peaked emission discussed in Section~\ref{sec:angular_results}. Support for ACE LAW 61 (correlated energy--angle distributions) in photonuclear data was introduced only in MCNP6.3.0~\cite{app_lanl_angular}, in anticipation of ENDF/B-VIII.1, and is therefore not relevant to the ENDF7u-based comparison performed here.
MCNPX 2.7.0 and MCNP6.2 are therefore expected to yield essentially identical photonuclear transport behavior whenever the same evaluated data are used, independent of the broader code developments introduced between the two versions (e.g., charged-particle transport, unstructured mesh, or fission-multiplicity sampling)~\cite{goorley2012mcnp6, app_mcnp62_release}.

To verify this expectation directly, all six broomstick benchmarks described in Section~\ref{sec:methods} were repeated with MCNP6.2, using the same ENDF7u photonuclear data, target geometry, and photon source definitions as the MCNPX calculations of Section~\ref{sec:results}. Two complementary comparisons were performed: the energy-resolved neutron yield, $\text{Y}_n$, defined consistently with Section~\ref{sec:methods} as the bin-integrated neutron current per source photon (n\,$\upgamma^{-1}$\,bin$^{-1}$), and the angular distribution of the escaping neutron current, expressed as the differential yield per unit solid angle, $\text{dY}_n/\text{d}\Omega$ (n\,$\upgamma^{-1}$\,sr$^{-1}$), both scored at the lateral target surface.

\begin{figure}[htbp]
    \centering
    \includegraphics[width=\textwidth]{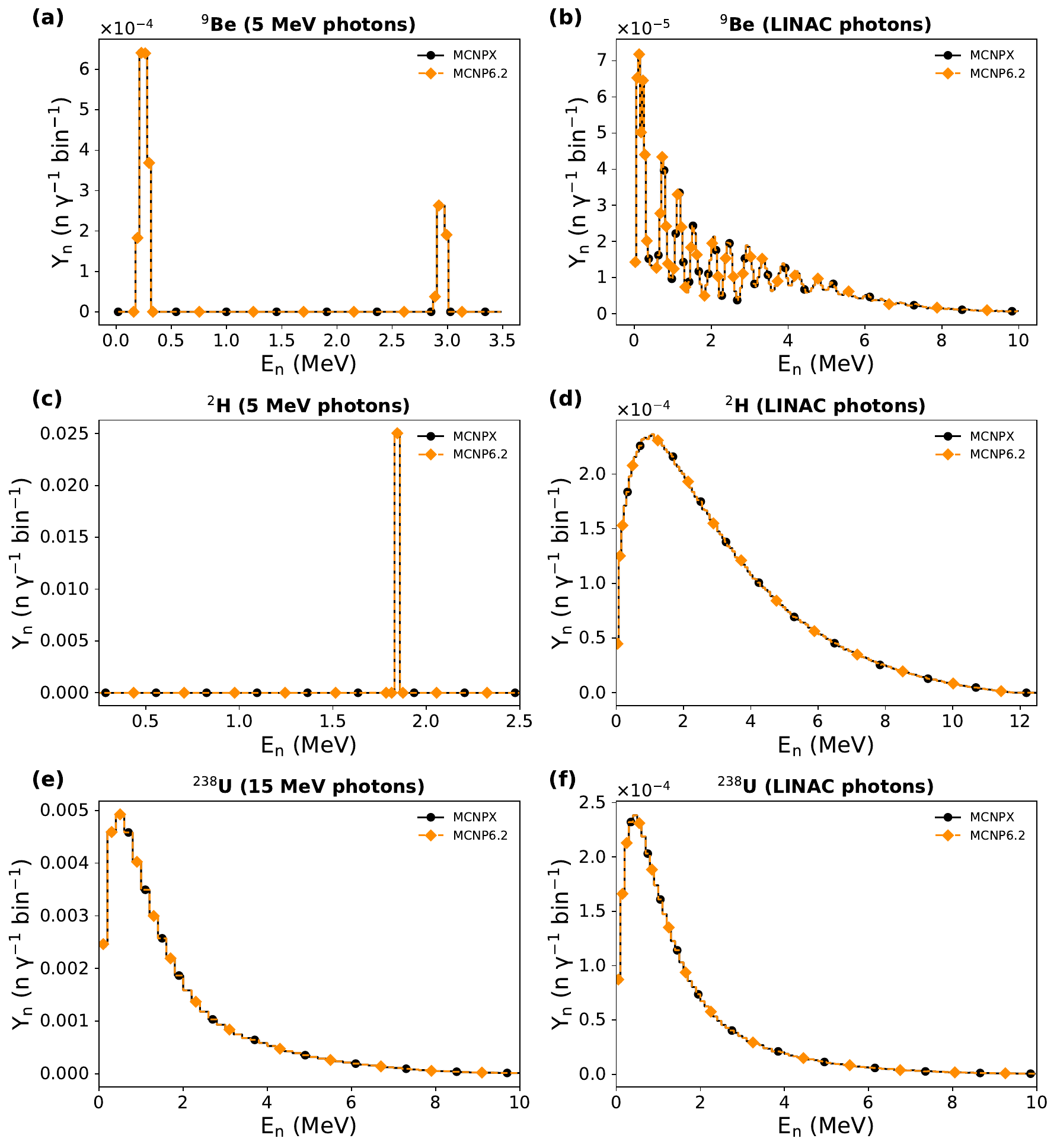}
    \caption{Energy-resolved neutron yield comparison between MCNPX and MCNP6.2
    for the six broomstick cases, using common ENDF7u photonuclear data.}
    \label{fig:app_energy}
\end{figure}

Figure~\ref{fig:app_energy} compares the energy-resolved neutron yields, $\text{Y}_n$, obtained with MCNPX and MCNP6.2 for all six configurations. The two codes are visually indistinguishable across the full energy range for all three target nuclides and both source types, consistent with the expectation that they sample the same tabulated reaction probabilities and secondary-energy distributions from the common ENDF7u evaluation.

\begin{figure}[htbp]
    \centering
    \includegraphics[width=\textwidth]{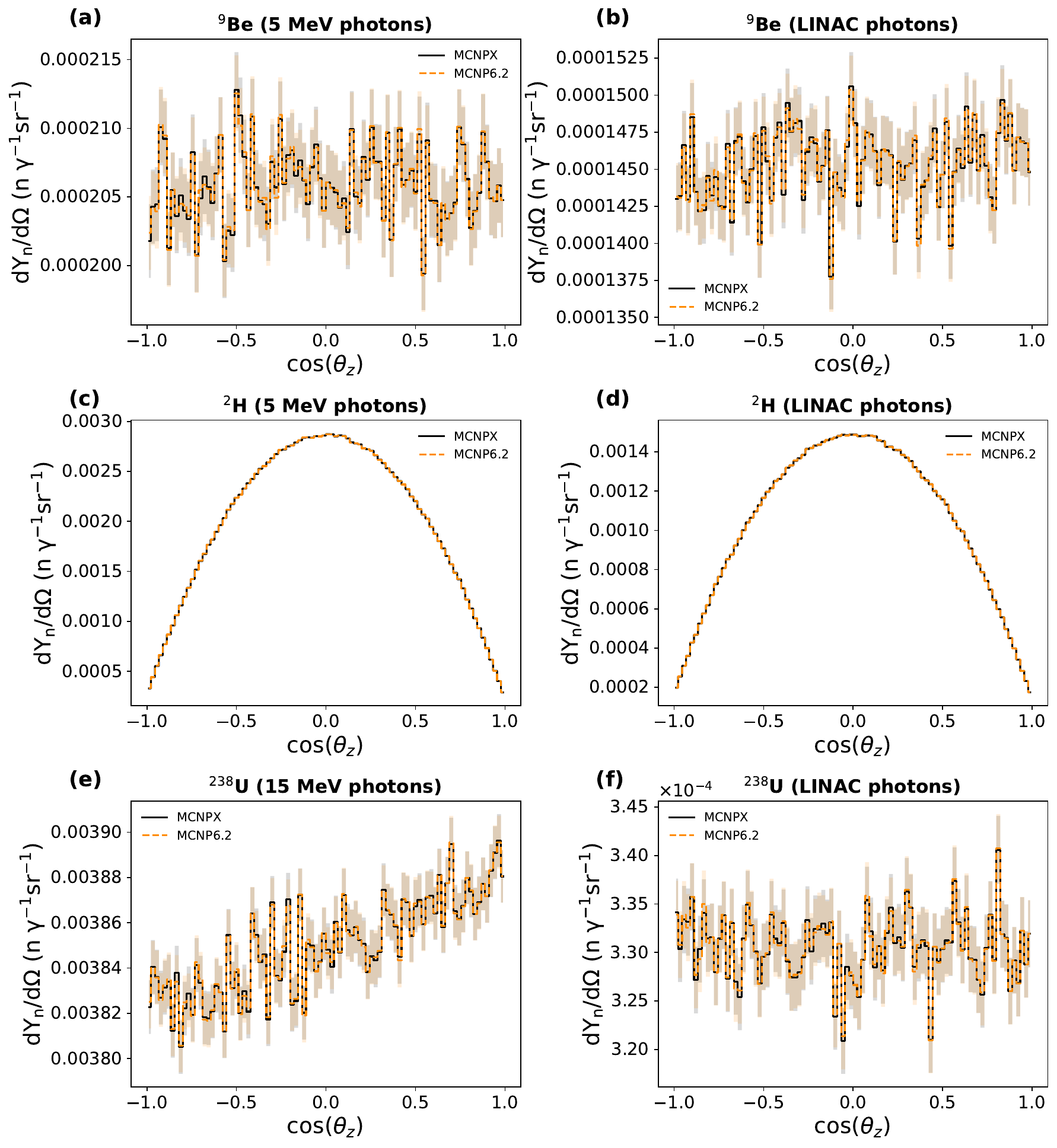}
    \caption{Angular distribution comparison (differential neutron current per unit solid
    angle versus cosine of polar angle) between MCNPX and MCNP6.2 for the six broomstick
    benchmarks, using common ENDF7u photonuclear data.}
    \label{fig:app_angle}
\end{figure}

Figure~\ref{fig:app_angle} compares the angular distributions of the escaping neutron current, $\text{dY}_n/\text{d}\Omega$, as a function of the cosine of the polar angle relative to the incident photon direction. As anticipated from the shared treatment of the secondary distributions described above, both codes reproduce the same angular shape for every case: an essentially isotropic distribution for $^{9}$Be and $^{238}$U, and the expected side-peaked angular dependence for $^{2}$H arising from the two-body kinematics of the $^{2}$H($\upgamma$,n) reaction. The solid-angle-integrated yields, obtained from Eq.~\eqref{eq:ytot_dual}, agree to within approximately 0.01\% or better in all six cases (e.g., $4.83794\times10^{-2}$ for MCNPX versus $4.83792\times10^{-2}$ for MCNP6.2 in the $^{238}$U 15-MeV case); the MCNPX values coincide with those reported in Table~\ref{tab:ytot_openmc_mcnpx}. The residual bin-to-bin fluctuations visible in some panels, most notably for $^{9}$Be, are consistent with the statistical uncertainty
of the Monte Carlo tallies rather than with a systematic difference between the two codes.

Taken together, these results confirm that MCNPX 2.7.0 and MCNP6.2 produce equivalent photonuclear transport results for the present verification when the same ENDF7u evaluated data are used. This supports the use of MCNPX 2.7.0 as the reference evaluated-data code throughout this work and indicates that the OpenMC verification
conclusions of Section~\ref{sec:results} are not contingent on the specific MCNP code version adopted for the reference calculations.

\pagebreak
\section*{Acknowledgments}
This research was supported in part through computational resources and services provided by Advanced Research Computing at the University of Michigan, Ann Arbor. This research received no specific grant from funding agencies in the public, commercial, or not-for-profit sectors.

\pagebreak
\bibliographystyle{style/ans_js}                              %custom ANS journal submission template bibliography style
\bibliography{bibliography}

\end{document}